\documentclass[prd,onecolumn,notitlepage,nofootinbib,superscriptaddress,showpacs,showkeys]{revtex4-1}

\usepackage{amsfonts}
\usepackage{graphicx}
\usepackage{dcolumn}
\usepackage{amsmath}
\usepackage{amssymb}
\usepackage[T1]{fontenc}
\usepackage[latin9]{inputenc}
\usepackage{esint}
\usepackage[brazil,english]{babel}
\usepackage[usenames,dvipsnames]{color}
\usepackage{longtable}
\usepackage{ulem}
\usepackage{nicefrac}
\usepackage{bm}
\usepackage{tikz}
\usepackage{braket}
\usepackage{graphicx, nicefrac}

\begin{document}

\title{Generating Kerr-anti--de Sitter thermodynamics}

\author{T. L. Campos}
\email{thiagocampos@usp.br}
\affiliation{Universidade de S\~{a}o Paulo, Instituto de F\'{\i}sica, Caixa Postal 66318,
05315-970, S\~{a}o Paulo-SP, Brazil}

\author{M. C. Baldiotti}
\email{baldiotti@uel.br}
\affiliation{Departamento de F\'{\i}sica, Universidade Estadual de Londrina, CEP
86051-990, Londrina-PR, Brazil}

\author{C. Molina}
\email{cmolina@usp.br}
\affiliation{Universidade de S\~{a}o Paulo, Escola de Artes, Ci\^{e}ncias e Humanidades, Avenida Arlindo Bettio 1000, CEP 03828-000, S\~{a}o Paulo-SP, Brazil}

\begin{abstract}
In the present work we study the construction of different thermodynamic descriptions for the Kerr-anti--de Sitter (KadS) black holes. The early versions of the KadS thermodynamics are briefly discussed, highlighting some of its strong points and shortcomings. Isohomogeneous transformations, a procedure for generating new thermodynamics, are presented and geometrically interpreted for KadS. This tool is used to determine possible KadS thermodynamics that can be constructed to satisfy a Smarr formula, and the validity of the first law in the generated thermodynamics. The connection between new thermodynamic theories and early Hawking's approach is considered. In this new framework, the usual KadS thermodynamics is complemented with its geometric construction, and Hawking's proposal, which does not satisfy the first law, is improved to an alternative thermodynamic theory. With the quantum statistical relation, Hawking's and this alternative KadS thermodynamics are also generalized from four to higher dimensions.
\end{abstract}

\maketitle

\section{Introduction}

Anti--de Sitter / field theory correspondences \cite{Maldacena:1997re,Witten:1998qj,Gubser:1998bc} play an important role in the investigation of fundamental aspects of the quantum properties of gravity. It is also a practical approach to gravity problems, providing computational tools in regimes that would be otherwise inaccessible with more traditional methods. In particular, rotating black holes have been studied in this context \cite{Guica:2008mu,Amado:2017kao}. One area where anti--de Sitter / field theory correspondences are being used is in the study of thermodynamic properties of black holes. From a theoretical point of view, black-hole thermodynamics is very interesting because it connects different branches of physics---general relativity, quantum mechanics and thermodynamics. Since thermodynamics is an effective theory of quantum statistics, it could serve as a guide to a deeper understanding of the quantum mechanical properties of gravity.

When the black hole is not asymptotically flat, additional difficulties are present. For instance, the Komar integral used to define the notion of energy diverges, leading to a generalized Komar mass. For Schwarzschild-anti--de Sitter, several routes have been used to study the associated thermodynamics \cite{Brown:1994gs,Louko:1996dw,Hemming:2007yq,kastor2009enthalpy,Elias:2018yct}. The addition of rotation to this picture, taking to the Kerr-anti--de Sitter (KadS) geometry, represents a further complication and it has been proposed in several works \cite{hawking1999rotation, gibbons2005first, caldarelli2000thermodynamics,dolan2011pressure,
dolan2012pdv,Cardoso:2013pza,Lemos:2015zma, kubizvnak2017black}.

In contrast to what occurs in traditional black-hole thermodynamics, several approaches to KadS can be found in the literature  \cite{hawking1999rotation,gibbons2005first, gao2023general}. On the other hand, the thermodynamics presented in \cite{caldarelli2000thermodynamics} is the most explored and it is often regarded as the thermodynamic theory for this spacetime. More recently, this field has gained interest due to Dolan's works \cite{dolan2011pressure, dolan2012pdv, dolan2010cosmological, dolan2011compressibility}. For instance, this author shows that this thermodynamics behaves similarly to a van der Waals gas. However, despite the extensive investigation of thermodynamic properties for this specific thermodynamic description of KadS, a proper geometric construction for this theory has not been sufficiently developed.

Another proposal for the thermodynamics of KadS can be extracted from \cite{hawking1999rotation}. Unlike the previous case, this is a geometric-oriented work, where thermodynamic quantities were defined via (renormalized) Komar integrals. However, as noted in \cite{gibbons2005first}, Hawking's proposal \cite{hawking1999rotation} does not satisfy a first law, and therefore it fails as a proper thermodynamic description. Alternatively, it was shown in \cite{gao2023general} that a generalization of the Iyer-Wald formalism \cite{iyer1994some}, which includes the cosmological constant, can lead to a different thermodynamics for KadS black holes. The goal of the present work is to discuss how these thermodynamic theories, and infinitely many others, can be
obtained.

A key feature of a thermodynamic theory is scale
invariance, which is implemented by requiring that all equations of state are given by homogeneous functions. Despite the peculiarities in black-hole
thermodynamics, where extensive and intensive properties are lost,
homogeneity still plays an important role \cite{kastor2009enthalpy,Baldiotti:2017ywq,Fontana:2018drk}. In the present work, 
explicit forms for transformations that preserve
homogeneity and the first law are given and multiple
thermodynamics can be obtained when they are applied to a given
thermodynamic description.

From a geometric point of view, although different Smarr formulas can be
obtained depending on a choice of a Killing field, there is no
guarantee that these thermodynamic variables will satisfy a first law. Nevertheless, the isohomogeneous transformations introduced here have a geometric
counterpart, which is implemented in the Killing vector fields that define
the Killing horizon. Thus, in this framework, we have the convergence of thermodynamic and geometric ideas. More precisely, the general thermodynamic approach involving isohomogeneous transformations is used to create Smarr formulas whose quantities, defined from Komar integrals, satisfy a first law. 

With this toolkit, a geometric construction for the usual KadS thermodynamics of \cite{caldarelli2000thermodynamics} can be given. Of all the possible KadS thermodynamics that can be generated from our procedure, one of them can be interpreted as the thermodynamic version of Hawking's proposal. Moreover, by combining this framework with the quantum statistical relation, Hawking's KadS thermodynamics is generalized to arbitrary dimensions.

The structure of this paper is presented as follows. In Sec.~\ref{sec:overview}, some main features of early treatments of the Kerr-anti--de
Sitter thermodynamics are reviewed, focusing on Hawking's and the usual proposals. In Secs.~\ref{sec: smarr transf} and \ref{sec: smarr}, isohomogeneous transformations are presented and interpreted
geometrically. These are the main tools developed in this paper for the
thermodynamic analysis of Kerr-anti--de Sitter black holes. Applications of
the isohomogeneous transformations are presented in Sec.~\ref{sec:
applying}. In particular, Hawking's and the usual
models are discussed and an alternative Kerr-anti--de
Sitter thermodynamics is constructed. In Sec.~\ref{sec:extensions},
extensions of the presented results to arbitrary dimensions are derived.
Final comments are presented in Sec.~\ref{remarks}.
In this paper, we use the geometric unit system and signature $(-,+,+,+)$.

\section{Overview of KadS thermodynamics}
\label{sec:overview}

\subsection{Four-dimensional Kerr-anti--de Sitter spacetime}

Kerr-anti--de Sitter spacetime is a stationary and axisymmetric spacetime
which models an asymptotically anti--de Sitter spinning black hole. A given
KadS black hole is specified by a choice of the mass parameter $m$,
rotation parameter $a$ and (negative) cosmological constant $\Lambda$. The
set $\{(m,a,\Lambda)\}$ parametrizes all possible four-dimensional KadS
solutions. In Boyer-Lindquist-like coordinates, the line element for this
geometry is written as \cite{griffiths2009exact} 
\begin{equation}
\mathrm{d}s^{2} = -\frac{\Delta _{r}}{\rho ^{2}} \left( \mathrm{d}t - \frac{%
a\sin^{2}\theta \, \mathrm{d} \phi }{\Xi}\right)^{2} + \frac{\Delta_{\theta}
\sin^{2}\theta }{\rho ^{2}} \left[ a \, \mathrm{d}t - \frac{\left(
r^{2}+a^{2} \right) \, \mathrm{d}\phi }{\Xi }\right]^{2} + \frac{\rho^{2}}{%
\Delta_{r}} \, \mathrm{d}r^{2} + \frac{\rho^{2}}{\Delta_{\theta }} \, 
\mathrm{d}\theta^{2} \, ,  \label{Boyer-Lindquist Kerr-anti de Sitter}
\end{equation}
with
\begin{align}
\begin{split}
& \Delta _{r}\equiv \frac{L^{2}+r^{2}}{L^{2}}\left( r^{2}+a^{2}\right)
-2mr\,,~\Delta _{\theta }\equiv 1-\frac{a^{2}}{L^{2}}\cos ^{2}\theta \ , \\
& \rho ^{2}\equiv r^{2}+a^{2}\cos ^{2}\theta \,,~\Xi \equiv 1-\frac{a^{2}} {%
L^{2}}\,,~L^{2}\equiv -\frac{3}{\Lambda }\,.
\end{split}
\label{quantidades BL1}
\end{align}
In this coordinate system, the coordinate $t$ was chosen such that
stationarity is implemented by the Killing vector field $\partial_t$, and
axial symmetry is expressed by the Killing vector field $\partial_\phi\,$.
We are interested in the nonextreme regime, where the spacetime has two
Killing horizons with topology $S^{2} \times \mathbb{R}$. The external
horizon is located at $r=r_{+}$, where $r_{+}$ is the largest positive real
root of the function $\Delta_{r}$. This horizon is considered as the
boundary of the KadS black hole. To enforce the Lorentzian character of
the metric, a necessary condition for the validity of the chart $%
(t,r,\theta,\phi)$ is $\Xi>0$.

The line element~\eqref{Boyer-Lindquist Kerr-anti de Sitter} can have its components rearranged to the following useful form, from which the angular velocity of the black hole can be easily extracted, 
\begin{equation}
\mathrm{d}s^{2} = -N^{2} \, \mathrm{d}t^{2} + \frac{\rho ^{2}}{\Delta _{r}}
\, \mathrm{d}r^{2} + \frac{\rho ^{2}}{\Delta _{\theta }} \, \mathrm{d}\theta
^{2} + \frac{\Sigma^{2} \sin^{2}\theta }{\rho ^{2} \, \Xi^{2}} \left( 
\mathrm{d}\phi - \omega \, \mathrm{d}t\right) ^{2} \, .
\label{second form KadS}
\end{equation}
The functions $\Sigma$, $N$, and $\omega$ are defined as 
\begin{align}
\begin{split}
& \Sigma ^{2} \equiv \left( r^{2}+a^{2} \right)^{2}\Delta _{\theta } -
a^{2}\Delta_{r}\sin^{2}\theta \, , \,\, N^{2} \equiv \frac{\rho ^{2}\Delta
_{r}\Delta _{\theta }}{\Sigma ^{2}} \, , \\
& \omega \equiv \frac{a\Xi }{\Sigma ^{2}} \left[\Delta _{\theta
}(r^{2}+a^{2})-\Delta _{r} \right] \, .
\end{split}
\label{quantidades BL2}
\end{align}
The quantity $\omega $ can be interpreted as the angular velocity for the
so-called zero-angular-momentum observer with respect to the
coordinate system $(t,r,\theta,\phi)$. Moreover, it has the following limits:
\begin{equation}
\lim_{r\rightarrow r_{+}}\omega =\Omega _{H} \ , \ \lim_{r\rightarrow \infty
}\omega =-\frac{a}{L^{2}}~,
\end{equation}
where 
\begin{equation}
\Omega_{H} \equiv \frac{a\,\Xi}{a^{2}+r_{+}^{2}}
\end{equation}
is interpreted as the angular velocity of the black hole and $-\nicefrac{a}{L^2}$ is the angular velocity of the spacetime at infinity,
with respect to the chart $(t,r,\theta,\phi)$ \cite{caldarelli2000thermodynamics}.

The surface $r=r_{+}$ is a Killing horizon for a Killing field of form 
\begin{equation}
K^{\mu } \equiv \xi ^{\mu } + \Omega \varphi ^{\mu } \, ,  \label{killing}
\end{equation}
where $\xi^\mu$ and $\varphi^\mu$ express stationarity and axial symmetry
respectively, and $\Omega$ is a scalar. The simplest case is 
\begin{equation}
\xi^\mu = \partial_t \ , \ \varphi^\mu = \partial_\phi~ \ , \ \Omega =
\Omega_H \, .  \label{coordinate basis}
\end{equation}
With these choices, the surface gravity associated to this horizon is given
by 
\begin{equation}
\kappa_{+} = \frac{\left( L^{2}+3r_{+}^{2} \right) r_{+}^{2} - a^{2} \left(
L^{2} - r_{+}^{2} \right)} {2L^{2}r_{+} \left( r_{+}^{2} + a^{2} \right)} \,
,  \label{eq:kappa}
\end{equation}
and the area of its two-dimensional spatial sections is 
\begin{equation}
A = \frac{4\pi \left(r_{+}^{2} + a^{2} \right)}{\Xi} \, .
\end{equation}
Moreover, it is useful to express the mass parameter as 
\begin{equation}
m = \frac{\left( r_+^2 + a^2 \right) \left(L^2 + r_+^2\right)}{2 r_+ L^2} \,
.
\end{equation}

\subsection{Early KadS thermodynamics}

\label{early-thermodynamics}

One of the first works to consider the thermodynamics of the KadS
geometry is due to Hawking \textit{et al.} \cite{hawking1999rotation}. In
the thermodynamic treatment of KadS black holes, a nonextreme KadS
spacetime (for a given choice of the parameters $m$, $a$, and $\Lambda$) is
associated with a thermal-equilibrium state. The set of possible
(nonextreme) KadS spacetimes forms the thermodynamic ensemble. It is
assumed in \cite{hawking1999rotation} that the mass ($M_{H}$) and the
angular momentum ($J$), 
\begin{equation}
M_{H} \equiv \frac{m}{\Xi }\, , \,\,\, J \equiv \frac{am}{\Xi ^{2}} \, ,
\label{Hawking definitions for m and J}
\end{equation}
are given by generalized Komar integrals associated, respectively, to the
Killing vector fields in Eq.~\eqref{coordinate basis}. In the present work,
the subscript $H$ refers to Hawking's proposal.

One important characteristic of the treatment based on the quantities $M_{H}$ and $J$ is that there is an associated Smarr formula: 
\begin{equation}
M_{H} = 2TS + 2\Omega_{H} J - 2V_{H} P \, ,
\label{smarr-hawking}
\end{equation}
where 
\begin{align}
\begin{split}
& T \equiv \frac{\kappa _{+}}{2\pi } =\frac{%
\left(L^{2}+3r_{+}^{2})r_{+}^{2}-a^{2}(L^{2}-r_{+}^{2} \right)} {4\pi
L^{2}r_{+}(r_{+}^{2}+a^{2})} \, , \,\, P \equiv -\frac{\Lambda}{8 \pi} = 
\frac{3}{8\pi L^{2}} \, , \\
& S \equiv \frac{A}{4} =\pi \frac{r_{+}^{2}+a^{2}}{\Xi } \, , \,\, V_{H}=%
\frac{4\pi }{3} \frac{r_{+}(r_{+}^{2}+a^{2})}{\Xi } \, .
\end{split}
\label{hawking-thermodynamics}
\end{align}

The notions of temperature and entropy in Eq.~\eqref{hawking-thermodynamics}
coincide with the standard view of usual black-hole thermodynamics \cite%
{bardeen1973four}. The pressure term $P$ can be understood by interpreting
the cosmological constant as a (dark-) energy contribution in the Universe 
\cite{ryden2017introduction}, with a constant energy density $\rho \equiv %
\nicefrac{\Lambda}{8\pi}$ and equation of state $\rho = - P$. This is a
typical viewpoint in cosmology. The volume term is better understood in the
pseudo-Cartesian coordinates of the Kerr-Schild form for the metric.
Surfaces of constant $r$ ($r=r_{+}$) are ellipsoids with Euclidean volume $%
V_{H}$ in Eq.~\eqref{hawking-thermodynamics} \cite{ballik2013vector}.

Relation~\eqref{smarr-hawking} is a version of the Smarr formula for nonzero cosmological constant. Considering a Killing field normal to the
horizon expressed as Eq.~\eqref{killing}, we propose writing the generalized
Smarr formula as 
\begin{align}
\begin{split}
\underbrace{-\frac{1}{8\pi }\int \nabla _{\mu }\xi _{\nu }\mathrm{d}A^{\mu
\nu }+\frac{\Lambda }{4\pi }\int g_{\mu \nu }\xi ^{\mu }\mathrm{d}\Sigma
^{\nu }}_{M}& -\underbrace{2\Omega \frac{1}{16\pi }\int \nabla _{\mu
}\varphi _{\nu }\mathrm{d}A^{\mu \nu }}_{2\Omega J} \\
& =\underbrace{-\frac{1}{8\pi }\int \nabla _{\mu }K_{\nu }\mathrm{d}A^{\mu
\nu }}_{2TS}+\underbrace{\frac{\Lambda }{4\pi }\int g_{\mu \nu }\xi ^{\mu }%
\mathrm{d}\Sigma ^{\nu }}_{-2V P} \, ,
\end{split} 
\label{gen smarr}
\end{align}
where $A$ is a spatial two-surface at the horizon, and the vector volume is used to integrate over a hypersurface $\Sigma$ extending from the singularity to $r=r_{+}$. The vector volume is the integral over a hypersurface of a divergence-free vector field $\xi^\mu$.%
\footnote{This is an operational definition, which is formalized in \cite{ballik2013vector}.}
Adopting~\eqref{coordinate basis}, Eq.~\eqref{gen smarr} reproduces Hawking's Smarr formula~\eqref{smarr-hawking}.

In Eq.~\eqref{gen smarr}, the quantity $M$ is a (modified) Komar mass,
\begin{equation}
M = -\frac{1}{8\pi }\int \nabla _{\mu }\xi _{\nu }\mathrm{d}A^{\mu
\nu }+\frac{\Lambda }{4\pi }\int g_{\mu \nu }\xi ^{\mu }\mathrm{d}\Sigma
^{\nu } \, .
\label{eq:KomarMass}
\end{equation}
The second term on the right-hand side of Eq.~{\eqref{eq:KomarMass}} can be interpreted as a contribution of the cosmological constant to $M$. If the integration were to be extended from the horizon to spatial infinity, as is usually performed for asymptotically flat spacetimes \cite{poisson2004relativist}, the quantity $M$ would be the usual Komar mass, which would diverge for $\Lambda \ne 0$ \cite{kastor2009enthalpy, hawking1999rotation}. In this case, a possible approach would be to renormalize the $\Lambda$ divergence, by subtracting an appropriately chosen background. However, such a procedure is generally not unique, since the choice of the reference background may be ambiguous \cite{caldarelli2000thermodynamics}.

In the present work, a quasilocal approach was implemented in~\eqref{gen smarr} using the concept of vector volume. With this method,
the generalized Smarr formula emerges in a straightforward way. More precisely, the traditional Smarr formula is based on the fact that, for a Killing vector $K^\mu$, the term $\nabla_\mu \nabla_\nu K^\mu$ vanishes for vacuum solutions of the Einstein gravitational field equation. This result is used to generate conserved charges from Stokes' theorem. However, in general cases, the vanishing object is $\nabla_\mu \nabla_\nu K^\mu - R_{\nu\mu} K^{\mu} \,$. In the absence of a conventional energy-momentum tensor but with a cosmological constant, $R_{\mu\nu} = \Lambda g_{\mu\nu}$. This justifies the extra terms that appear in Eq.~\eqref{gen smarr}.%
\footnote{It was also used the property that $\varphi^\mu$ does not contribute to the vector volume (the integral vanishes for it) and, thus, the vector volume
for $K^\mu$ is the same as the one for $\xi^\mu$.}

It should be stressed that in Hawking's work \cite{hawking1999rotation}, no
consideration about the first law was made. However, as noted in \cite{gibbons2005first}, it can be verified that%
\footnote{In \cite{gibbons2005first}, the authors do not treat the cosmological
constant as a thermodynamic variable. However, even if $\Lambda$ is
considered a thermodynamic pressure, the inequality is not corrected. In
this case, one would still obtain $\text{d} M_{H} \neq T \, \text{d} S +
\Omega_{H} \, \text{d} J + V_{H} \, \text{d} P \,$.}
\begin{equation}
\mathrm{d}M_{H} \neq T \, \mathrm{d}S + \Omega _{H} \, \mathrm{d}J \, .
\label{Hawking-problem}
\end{equation}
From Eq.~\eqref{Hawking-problem}, we come to an important conclusion: in Hawking's proposal for a KadS thermodynamics, the quantity $M_{H}$ can not be associated to a well-defined notion of free energy.

In order to construct a thermodynamic theory based on KadS spacetime
which is consistent with the first law, Caldarelli \textit{et al.} \cite%
{caldarelli2000thermodynamics} and Gibbons \textit{et al.} \cite%
{gibbons2005first} considered modified definitions for energy and angular
momentum, 
\begin{equation}
M_{U} \equiv \frac{m}{\Xi ^{2}} \, , \,\,\, J \equiv \frac{am}{\Xi ^{2}} \, ,
\end{equation}
connected to generalized Komar integrals associated to the Killing fields 
\begin{equation}
\frac{\partial_t}{\Xi} \,\, \text{and} \,\, \partial_\phi \,,
\end{equation}
respectively. In the present work, the subscript $U$ refers to usual thermodynamic theory, abbreviated UTT.

This approach attracted further attention with the contribution of Dolan 
\cite{dolan2011pressure, dolan2012pdv}. In those works, an association of $\Lambda $ with pressure is made and the first law is written as 
\begin{equation}
\mathrm{d}M_{U}=T\mathrm{d}S+\Omega _{U}\mathrm{d}J+V_{U}\mathrm{d}P~,
\end{equation}
where 
\begin{align}
& \Omega _{U} \equiv \Omega _{H}+\frac{a}{L^{2}} = \frac{a}{a^{2}+r_{+}^{2}}
\left(1+\frac{r_{+}^{2}}{L^{2}}\right) \, , \\
& V_{U} \equiv V_{H}+\frac{4\pi }{3}a^{2}M_{U} =\frac{2\pi }{3} \frac{\left(
r_{+}^{2} + a^{2}\right) \left(2r_{+}^{2}L^{2} +
a^{2}L^{2}-r_{+}^{2}a^{2}\right)} {L^{2}\Xi^{2}r_{+}} \, .
\end{align}
It can be checked that, with these choices for the thermodynamic variables,
Smarr formula is also satisfied. However, a relevant point to be addressed
in the present work is that, with $\Omega = \Omega_U$, it is obtained from Eq.~\eqref{killing} 
\begin{equation}
\xi^\mu \neq \frac{\partial_t}{\Xi} \, , \,\, \varphi^\mu = \partial_\phi \,
,
\end{equation}
which is different from what one would expect. The correct form for $K^\mu$ will be obtained from our formalism and presented in  Sec.~\ref{sec: applying}.

\subsection{Homogeneity and Euler relation}

Homogeneity of the equations of state is a crucial feature in black-hole thermodynamics \cite{kastor2009enthalpy,Baldiotti:2017ywq}. 
%
%
A function $M\left( X_{1},X_{2},X_{3},\ldots \right)$ is said to be a homogeneous function of degree $r$, on the variables $X_{i}$ of degree $\alpha _{i}$, if
\begin{equation}
M\left( \lambda ^{\alpha _{1}}X_{1},\lambda ^{{\alpha _{2}}}X_{2},\lambda
^{\alpha _{3}}X_{3},\ldots \right) =\lambda ^{r}M_{0}\left(
X_{1},X_{2},X_{3},\ldots \right) \, .
\end{equation}
For such functions, Euler's theorem for homogeneous functions states that 
\begin{equation}
rM=\alpha _{1}X_{1}\frac{\partial M}{\partial X_{1}}+\alpha _{2}X_{2}\frac{
\partial M}{\partial X_{2}}+\alpha _{3}X_{3}\frac{\partial M}{\partial X_{3}}
+\ldots \,.  \label{eq:euler-theorem}
\end{equation}
In the thermodynamic-oriented literature, Eq.~\eqref{eq:euler-theorem} is
known as Euler relation. In early versions of KadS thermodynamics
(discussed in Sec.~\ref{early-thermodynamics}), the independent variables
are $\{ S,J,P \}$, function $M$ denotes the black-hole mass, and we have $\alpha _{1}=\alpha _{2}=-\alpha _{3}=2$ and $r=1$. One can read from the
dimensions of these quantities: 
\begin{equation}
\left[ S\right] \sim \left[ J\right] \sim [\mathrm{length}]^{2} \, , \,\, %
\left[ P\right] \sim [\mathrm{length}]^{-2} \,, \,\, \left[ M\right] \sim [%
\mathrm{length}] \, .
\end{equation}
A scaling argument, based on the relationship between dimension and
homogeneity, is used in \cite{kastor2009enthalpy} to justify the inclusion
of $P \equiv -\nicefrac{\Lambda}{8\pi}$ as a pressure in the thermodynamic
description. This relation can be employed, by directed comparison of Smarr and Euler relations, to determine the equations of states of the thermodynamic description \cite{gauntlett1999black,townsend2001first}.

However, this identification method should be applied with caution. The connection between dimensional analysis and homogeneity can lead to the conclusion that the Smarr relation is always the geometric counterpart of the Euler relation. But not all Smarr relations can be identified as Euler relations. A concrete example is given by Eq.~\eqref{smarr-hawking}. Although it is a proper Smarr relation, it is not an Euler relation because it does not provide a first law. This means that this procedure is not always reliable to furnish proper thermodynamic descriptions. Tools for a more careful analysis will be developed in the next section.

\section{Isohomogeneous transformations}

\label{sec: smarr transf}

In the present work we study the construction of different thermodynamic
descriptions for the KadS black hole. Let us motivate the main ideas
used in this section. It is believed that the Smarr relation represents the
geometric side of the Euler relation for homogeneous functions \cite{kastor2009enthalpy}. However, as discussed in Sec.~\ref{early-thermodynamics}, Hawking's proposal shows that this is not true in general. Indeed, from Eq.~\eqref{smarr-hawking}, 
\begin{equation}
M_{H} = 2TS + 2\Omega_{H}J - 2 V_{H} P \neq 2\frac{\partial M_{H}}{\partial S%
}S+2\frac{\partial M_{H}}{\partial J}J-2\frac{\partial M_{H}}{\partial P}P
\, .
\end{equation}
In this section we analyze under which circumstances a Smarr relation does
not represent an Euler relation. 
We develop tools that allow to transform this Smarr relation into a valid Euler relation that gives a proper first law. Furthermore, the geometrical significance of this thermodynamic development is clarified in the next section.

Let us consider a thermodynamic potential $M_{0}=M_{0}\left(S,X_{2},X_{3},\ldots \right)$, that is, a homogeneous function of degree $r$ such that 
\begin{equation}
M_{0}\left( \lambda ^{\alpha _{1}}S, \lambda ^{\alpha _{2}}X_{2},
\lambda^{\alpha _{3}}X_{3},\ldots \right) = \lambda ^{r}M_{0}\left(
S,X_{2},X_{3},\ldots \right) \, ,
\end{equation}
\begin{equation}
\mathrm{d}M_{0} = T_{0} \, \mathrm{d}S + \sum_{i} Y_{0}^{i} \, \mathrm{d}%
X_{i} \, ,  \label{eq_dM}
\end{equation}
where $\alpha_{i}$ is the degree of $i$th independent variable. Since $M_{0}$ is a homogeneous function, it obeys Euler relation: 
\begin{equation}
rM_{0}=\alpha _{1}ST_{0}+\alpha _{2}X_{2}Y_{0}^{2}+\alpha
_{3}X_{3}Y_{0}^{3}+\cdots \, .  \label{oer}
\end{equation}

Our goal is to analyze the transformations on thermodynamic potential $M_{0}=M_{0}\left(S,X_{2},X_{3},\cdots \right) $ that preserve homogeneity.
Consider a homogeneous function $g=g\left( S,X_{2},X_{3},\cdots \right)$ of
degree zero, 
\begin{equation}
\alpha _{1}S\frac{\partial g}{\partial S}+\alpha _{2}X_{2}\frac{\partial g}{
\partial X_{2}}+\alpha _{3}X_{3}\frac{\partial g}{\partial X_{3}}+\cdots =0
\, .  \label{oer2}
\end{equation}
Let us also assume that $g$ is positive definite. Multiplying $rM_{0}$ by $g$
and using Eq.~\eqref{oer}, 
\begin{equation}
rgM_{0}=\alpha _{1}gST_{0}+\alpha _{2}gX_{2}Y_{0}^{2}+\alpha_{3}gX_{3}Y_{0}^{3}+\cdots \, .  \label{er}
\end{equation}
Adding expression~\eqref{oer2} to Eq.~\eqref{er}, 
\begin{equation}
rgM_{0} = \alpha_{1} g S T_{0} + \alpha _{2}gX_{2}Y_{0}^{2} + \cdots +
h\left( \alpha_{1}S\frac{\partial g}{\partial S} + \alpha_{2}X_{2}\frac{%
\partial g}{\partial X_{2}}+\cdots \right) \, ,  \label{Eq-rgM0}
\end{equation}
for an arbitrary function $h$.

To preserve homogeneity, we impose that $h$ is a general homogeneous
function with the same degree as $M_{0}$. Thus, from the original Euler relation~\eqref{oer}, it follows that
\begin{equation}
rM_{1} = \alpha _{1}ST_{1} + \alpha _{2}X_{2}Y_{1}^{2} +
\alpha_{3}X_{3}Y_{1}^{3}+\cdots \,,  \label{rl}
\end{equation}
where 
\begin{equation}
M_{1} \equiv g M_{0} \, , \,\, T_{1} \equiv gT_{0}+h\frac{\partial g}{%
\partial S} \, , \,\, Y_{1}^{i} \equiv gY_{0}^{i}+h\frac{\partial g}{%
\partial X_{i}} \,.  \label{sg}
\end{equation}
Relation~\eqref{rl} is equivalent to Eq.~\eqref{oer}.

For any positive definite homogeneous function $g$ of degree zero and for
any homogeneous function $h$ of degree $r$, we call Eq.~\eqref{sg} an isohomogeneous transformation. The next step is to determine when Eq.~\eqref{rl} is also an Euler relation.

But while relation~\eqref{rl} enforces homogeneity, not all functions $M_{1}$
can be identified as a legitimate thermodynamic potential. Specifically, from Eq.~\eqref{sg}, 
\begin{equation}
T_{1}\,\mathrm{d}S+\sum_{i}Y_{1}^{i}\,\mathrm{d}X_{i}\,\,\,=g\,\mathrm{d}%
M_{0}+h\,\mathrm{d}g=\mathrm{d}M_{1}+\left( h-\frac{M_{1}}{g}\right) \mathrm{%
d}g\,.  \label{fs}
\end{equation}
It follows from Eq.~\eqref{fs} that, for a nonconstant $g$, $M_{1}$ will be
a thermodynamic potential only if $h=\nicefrac{M_{1}}{g}=M_{0}$. In this
case, the right-hand side is an exact differential, 
\begin{equation}
h=M_{0}\Longrightarrow \mathrm{d}M_{1}=T_{1}\,\mathrm{d}S+\sum_{i}Y_{1}^{i}\,%
\mathrm{d}X_{i}\,,
\end{equation}%
and Eq.~\eqref{rl} is a valid Euler relation. However, as we will see, the thermodynamic descriptions associated to $M_{0}$ and $M_{1}$ are different.

When Eq.~\eqref{rl} is an Euler relation, satisfying a first law of thermodynamics, we call this relation a thermodynamic Smarr formula. An example of a thermodynamic Smarr formula
is the one from the UTT \cite{caldarelli2000thermodynamics}. The other
cases will be classified as nonthermodynamic Smarr formulas. For
instance, Hawking's proposal \cite{hawking1999rotation} (discussed in
Sec.~\ref{early-thermodynamics}) involves a nonthermodynamic Smarr
formula. Given a thermodynamic Smarr formula with a thermodynamic potential $M_{0}$, the transformation to another thermodynamic Smarr formula, which
follows from choosing a homogeneous function of degree zero $g$ and fixing $%
h=M_{0}$, will be called a exact isohomogeneous transformation.%
\footnote{Exact because the left-hand side of Eq.~\eqref{fs} is an
exact differential.}

One remark is in order. A given thermodynamic description is characterized by a set of equations of state $(T,\{Y^{i}\})$. In principle more than one description can share some specific equations of state. For instance, we can search for all thermodynamics that share the same temperature. In this example, given a legitimate thermodynamic
potential $M_{0}$ characterized by a temperature $T_{0}$, other legitimate
potentials $\{\tilde{M}\}$ are obtained setting $h=M_{0}$. In fact, imposing
that they are also characterized by the same $T_{0}\,$, 
\begin{equation}
\tilde{T}=T_{0} \Longrightarrow \left( \tilde{g}-1\right) T_{0} + M_{0} 
\frac{\partial \tilde{g}}{\partial S}=0 \, .  \label{eq-T0}
\end{equation}
To clarify, tilted versions are used to explicit the fact that the
temperature ($\tilde{T}$) and function ($\tilde{g}$) are not being fixed in
the development.

Multiple solutions of Eq.~\eqref{eq-T0} can be derived. Indeed, for any
arbitrary function $f$ (with degree $r$) that does not depends on $S$, we
obtain 
\begin{equation}
\tilde{g} = 1-\frac{f\left( X_{i}\right)}{M_{0}} \, .  \label{g-til}
\end{equation}
The presented development can be extended for several equations of state.
That is, given a thermodynamics with $N$ variables, it is possible to
construct another one preserving $(N-1)$ equations of state. In particular,
multiple thermodynamic descriptions (for adS black holes) can have the same
definition of temperature that is used by Hawking in \cite{hawking1999rotation}.

\section{Geometry of isohomogeneous transformations on KadS}
\label{sec: smarr}

In the previous section, general ideas involving isohomogeneous transformations in a thermodynamic context were presented. In the present section, this development will be discussed geometrically, considering the KadS spacetime. Adapting the nomenclature of Sec.~\ref{sec: smarr transf} to
the four-dimensional KadS black holes, we have 
\begin{align}
\begin{split}
& r=1 \, , \,\, \alpha _{1}=2 \, , \,\, \alpha _{2}=2 \, , \,\, \alpha
_{3}=-2 \, , \\
& X_{2}=J \, , \,\, X_{3}=P \, , \\
&Y_{0}^{2} = \Omega _{0} \, , \,\, Y_{0}^{3}=V_{0} \, , \,\,
Y_{1}^{2}=\Omega _{1} \, , \,\, Y_{1}^{3}=V_{1} \, .
\end{split}
\label{var}
\end{align}
The general expressions~\eqref{rl} and \eqref{sg} imply that, for the
KadS case, 
\begin{equation}
M_{1} = 2 ST_{1} + 2 J \Omega _{1} - 2 P V_{1} \,,  \label{rl-KadS}
\end{equation}
where 
\begin{equation}
M_{1} = g M_{0} \, , \,\, T_{1} = gT_{0} + h\frac{\partial g}{\partial S} \,
, \,\, \Omega_{1} = g \Omega_{0} + h\frac{\partial g}{\partial J} \, , \,\,
V_{1} = g V_{0} + h\frac{\partial g}{\partial P} \,.  \label{sg-KadS}
\end{equation}

The geometric counterpart to the isohomogeneous transformations is implemented on the Killing fields that define the Killing horizon. The resulting transformed object must also be a Killing field normal to the horizon, allowing its use as a generator of a new Smarr formula. By adopting this geometric perspective, the distinct thermodynamic theories for KadS can be understood within the framework of rotating reference frames.

In accordance with previous section, we introduce a positive function $g$ of the thermodynamic variables $\{ S, J, P \}$. With $g$, other Smarr formulas can be obtained, rescaling the Killing field in Eq.~\eqref{killing} as
\begin{equation}
K^{\mu} \longrightarrow gK^{\mu }=g\xi ^{\mu }+g\Omega _{0}\varphi ^{\mu }
\, .  \label{g K}
\end{equation}
In stationary asymptotically flat spacetimes, there is a choice of $g$ which
normalizes $\xi^\mu$ at infinity \cite{bardeen1973four}. However, a different $g$ can be used to normalize $\xi^\mu$ at a finite distance. This gives the Tolman redshift factor for the temperature \cite{santiago2018tolman}. Therefore, keeping $g$ arbitrary is useful, especially when dealing with a nonasymptotically flat geometry.

On the other hand, this does not represent the entirety of possible
transformations in a spacetime with axial symmetry. We can also consider
combinations between the Killing fields themselves, 
\begin{equation}
K^{\mu }\longrightarrow gK^{\mu }=\left( g\xi ^{\mu }-h\frac{\partial g}{%
\partial J}\varphi ^{\mu }\right) +\left( g\Omega _{0}+h\frac{\partial g}{%
\partial J}\right) \varphi ^{\mu }\,,  \label{killing1}
\end{equation}
where $h$ is an arbitrary function of $\{S,J,P\}$. Equation~\eqref{killing1} has
been structured in a way that is consistent with our developments in the previous section. Notably, the resultant angular velocity is expressed in the format found in Eq.~\eqref{sg-KadS}. Moreover, although the quantities proportional to $h$ sum to zero on the right-hand side of Eq.~\eqref{killing1}, in this form, $h$
is associated with the change to a new frame that rotates with angular velocity 
\begin{equation}
\frac{\text{d}\phi_{1}}{\text{d}t} = \frac{\text{d}\phi}{\text{d}t} +\frac{h%
}{g}\frac{\partial g}{\partial J}\,,  \label{dot-phi}
\end{equation}
where $(t,r,\theta,\phi)$ is the original coordinate system. Furthermore, $g$ represents a choice of normalization for the Killing vector normal to the horizon, which, as we will see, rescales some thermodynamic
quantities.

The transformation on the Killing field~\eqref{killing1} propagates to the
Komar formulas, allowing a reinterpretation of the thermodynamic quantities: 
\begin{equation}
\begin{split}
& \underbrace{-\frac{g}{8\pi }\int \nabla _{\mu }\xi _{\nu }\mathrm{d}A^{\mu
\nu }+\frac{g\Lambda }{4\pi }\int g_{\mu \nu }\xi ^{\mu }\mathrm{d}\Sigma
^{\nu }}_{M_{1}}+\underbrace{h\frac{\partial g}{\partial J}\frac{1}{8\pi }%
\int \nabla _{\mu }\varphi _{\nu }\mathrm{d}A^{\mu \nu }}_{2h\frac{\partial g%
}{\partial J}J} -\underbrace{2\bigg(g\Omega _{0}+h\frac{\partial g}{\partial
J}\bigg)\frac{1}{16\pi }\int \nabla _{\mu }\varphi _{\nu }\mathrm{d}A^{\mu
\nu }}_{2\Omega _{1}J} \\
& \hspace{3.5cm}=\underbrace{-\frac{g}{8\pi }\int \nabla _{\mu }K_{\nu
}\mathrm{d}A^{\mu \nu }}_{2gT_{0}S}+\underbrace{\frac{g\Lambda }{4\pi }\int
g_{\mu \nu }\xi ^{\mu }\mathrm{d}\Sigma ^{\nu }}_{-2gV_{0}P}\,,
\end{split}
\label{transformed geometric Smarr}
\end{equation}
where $M_{1}$ and $\Omega _{1}$ are presented in Eq.~\eqref{sg-KadS}.

To identify Eq.~\eqref{transformed geometric Smarr} as a Smarr relation, the
term $2h\frac{\partial g}{\partial J}J$ must be eliminated. This can be done
by imposing that $g$ is a homogeneous function of degree zero. Therefore,
using Eq.~\eqref{oer2}, 
\begin{equation}
J\frac{\partial g}{\partial J}=P\frac{\partial g}{\partial P}-S\frac{%
\partial g}{\partial S}\,,  \label{define g}
\end{equation}
and Eq.~\eqref{transformed geometric Smarr} furnishes 
\begin{equation}
M_{1}+2h\bigg(P\frac{\partial g}{\partial P}-S\frac{\partial g}{\partial S}%
\bigg)-2\Omega _{1}J=2gT_{0}S-2gV_{0}P\,.
\end{equation}
Factorizing $S$ and $P$, we see that the temperature and volume change
according to Eq.~\eqref{sg-KadS} and Smarr relation~\eqref{rl-KadS} is
recovered. From a geometrical point of view, expression~\eqref{define g}
links surface and hypersurface integrals: 
\begin{equation}
\int \nabla_{\mu }\varphi _{\nu }\mathrm{d}A^{\mu \nu }~,\ \int \nabla_{\mu
}K_{\nu }\mathrm{d}A^{\mu \nu }~,\ \int g_{\mu \nu }\xi ^{\mu }\mathrm{d}%
\Sigma ^{\nu } \, .
\end{equation}

The terms multiplying $h$ have a zero net contribution to the change of the
Killing field in Eq.~\eqref{killing1}, but, from Eq.~\eqref{sg-KadS}, it is
observed that this rotating frame changes how the energy is distributed into
heat ($T\mathrm{d}S$) and mechanical work ($\Omega \mathrm{d}J+V\mathrm{d}P$%
). In the geometric language, the change in the angular velocity could be
interpreted as a rotation between frames adapted to different thermodynamic
descriptions.

It is worth emphasizing that what defines distinct thermodynamics are not different choices of coordinates, but rather different Killing fields and their contribution to the Komar integrals. Nevertheless, adapted coordinates facilitate geometric and physical interpretations behind each theory and also how they relate to each other. For instance, transformation~\eqref{killing1} implies that, for a coordinate system $(t,r,\phi,\theta)$, where $\xi ^{\mu }$ and $\varphi ^{\mu }$ are vectors from the coordinate basis ($\partial _{t}$ and $\partial _{\phi }$), other coordinates $t_{1}$ and $\phi _{1}$ can be chosen so that
\begin{equation}
\partial _{t_{1}}=g\partial _{t}-h\frac{\partial g}{\partial J}\partial
_{\phi }\,,\,\,\partial _{\phi _{1}}=\partial _{\phi }\,.  \label{parcial-t1}
\end{equation}%
The respective coordinate change is 
\begin{equation}
t=g\,t_{1}\,,\,\,\phi =\phi _{1}-h\frac{\partial g}{\partial J}t_{1}\,.
\label{change of coordinates}
\end{equation}
In the adapted coordinates, the Killing field becomes 
\begin{equation}
gK=\partial _{t_{1}}+\Omega _{1}\partial _{\phi _{1}}\,, \label{eq-eK}
\end{equation}
and Eq.~\eqref{dot-phi} is immediately recovered.

The discussion involving thermodynamic and nonthermodynamic Smarr formulas,
in the context of the isohomogeneous transformations, reappears with the
geometric formalism. Indeed, although the transformations in the Killing
field~\eqref{killing1} transforms a Smarr formula into an equally valid one, it may destroy the thermodynamic description of the system, generating thermodynamic variables that do not satisfy the first law. More precisely, even if the description with the Killing field~\eqref{killing} is related to a legitimate thermodynamic theory, 
\begin{equation}
\mathrm{d}M_{0}=T_{0}\mathrm{d}S+\Omega _{0}\mathrm{d}J+V_{0}\mathrm{d}P \, ,
\end{equation}
it is not always true that its transformed version also satisfies the first law. That is, it may be the case that the thermodynamic description is invalidated: 
\begin{equation}
\mathrm{d}M_{1}\neq T_{1}\mathrm{d}S+\Omega _{1}\mathrm{d}J+V_{1}\mathrm{d}P
\, .
\end{equation}
Nonetheless, from Sec.~\ref{sec: smarr transf}, we know that the first
law is preserved if the transformed Smarr formula is obtained by an exact isohomogeous transformation, i.e. taking $h = M_0$.

\section{Applying the isohomogeneous transformations}

\label{sec: applying}

\subsection{Generating thermodynamic theories from Hawking's proposal}

\label{sec: generating}

In Hawking's proposal, discussed in Sec.~\ref{early-thermodynamics}, the internal energy $M_H$ is given by 
\begin{equation}
M_{1}=M_{H}=\frac{m}{\Xi } \, .  \label{eh}
\end{equation}
It is an example of a nonthermodynamic theory, in the sense that it does
not have a well-defined first law: 
\begin{equation}
dM_{H}\neq T\mathrm{d}S+\Omega _{H}\mathrm{d}J+V_{H}\mathrm{d}P \, .
\label{eq-dM-H}
\end{equation}

It is possible to turn the inequality~\eqref{eq-dM-H} into an equality by
adding an extra term on the left side of Eq.~\eqref{eq-dM-H}: 
\begin{equation}
T\mathrm{d}S+\Omega _{H}\mathrm{d}J+V_{H}\mathrm{d}P=dM_{H}-\frac{M_{H}}{%
2\Xi }\mathrm{d}\Xi \,.  \label{th1}
\end{equation}
Comparing Eq.~\eqref{th1} with Eq.~\eqref{fs}, we have 
\begin{equation}
\frac{M_{H}}{2\Xi }=\left( \frac{M_{H}}{g}-h\right) \frac{\mathrm{d}g}{%
\mathrm{d}\Xi }\,,  \label{condition h}
\end{equation}
where $g$ is written as a function of $\Xi $. The development~\eqref{oer2}
only required that $g$ be a function of $S,J$, and $P$, so the
representation $g=g(\Xi )$ is a constraint on the functional form of $g$.

We emphasize that there are an infinite number of possible proper thermodynamic theories related to Hawking's proposal. In fact, consider the following choice for $g$ and the associated function $h$:
\begin{equation}
g=\Xi^{n} \Longrightarrow h=\frac{M_{H}}{\Xi ^{n}}\left( 1-\frac{1}{2n}%
\right) \, ,  \label{g-x}
\end{equation}
with $n\neq 0$. The (multiple) proper thermodynamics generated (with subindex $0$ following the notation of Sec.~\ref{sec: smarr transf}), related to Hawking's thermodynamics, are
\begin{equation}
T=\xi T_{0} \, , \,\,\, \Omega _{H}=\xi \Omega _{0}-\left( \xi -\Xi
^{n}\right) \frac{M_{H}}{J\Xi ^{n}} \, , \,\,\, V_{H}=\xi V_{0}-\left( \xi
-\Xi ^{n}\right) \frac{M_{H}}{2P\Xi ^{n}} \, ,  \label{nht}
\end{equation}
where 
\begin{equation}
\xi =\frac{\Xi^{n}} {1-\left( 2n-1\right) \left( 1-\Xi \right)} \, , \,\,\,
\Xi (J,S,P)=1+\frac{3}{8PS+3}-\frac{3\left( 4\pi ^{2}J^{2}+S^{2}\right) } {%
12\pi ^{2}J^{2}+S^{2}(8PS+3)} \, .  \label{xi}
\end{equation}

Note from Eq.~\eqref{sg-KadS} that the temperature for a theory will be
proportional to its surface gravity only when $h=0$. This corresponds to $\xi = g = \sqrt{\Xi}$ in Eq.~\eqref{xi}. Given a proper thermodynamic description where the temperature coincides with its surface gravity, characterized by energy $M_{0}\,$, all other cases are obtained by taking $h=M_{0}$. Thus, with a suitable normalization, there is a unique thermodynamic theory for which the temperature coincides with its surface gravity.

Of all the possible KadS thermodynamics that can be generated with the
formalism in this work, two will receive special attention in the remainder
of this section. The first is the UTT, which has been extensively studied in the literature. The second is the one that minimally modifies Hawking's proposal, which will be analyzed on a deeper level. The latter is, in a sense, the natural extension of Hawking's proposal to a KadS thermodynamics.

\subsection{From Hawking's proposal to the usual thermodynamic theory}

\label{subsec: hawking to dolan}

Using result~\eqref{g-x}, consider the following configuration: 
\begin{equation}
n=1\implies g=\Xi \, , \,\,\, h=\frac{M_{H}}{2\Xi } \, , \,\,\, \xi =1 \,.
\label{hawking-dolan}
\end{equation}
In this case, a valid thermodynamic theory connected to Hawking's proposal
is the one with energy 
\begin{equation}
M_{0}=\frac{M_{1}}{g}=\frac{M_{H}}{\Xi }=\frac{m}{\Xi ^{2}}=M_{U} \, ,
\label{ud}
\end{equation}
which is the UTT (discussed in Sec.~\ref{early-thermodynamics}).

Note that Eq.~\eqref{hawking-dolan} is not an exact isohomogeneous transformation because it links nonthermodynamic and thermodynamic Smarr
formulas. For the remaining thermodynamic variables, from Eq.~\eqref{sg-KadS},
\begin{align}
& T_{H} = \Xi T_U + \frac{M_H}{2 \Xi} \frac{\partial \Xi}{\partial S}
\implies T_U = T_H =T\,, \\
& \Omega_H = \Xi \Omega_U + \frac{M_H}{2 \Xi} \frac{\partial \Xi}{\partial J}
\implies \Omega _{U}=\Omega _{H}+\frac{a}{L^{2}}\,, \\
&V_H = \Xi V_U + \frac{M_H}{2 \Xi} \frac{\partial \Xi}{\partial P} \implies
V_{U}=V_{H}+\frac{4\pi }{3} a^{2}M_{U}\,.  \label{Dolan volume}
\end{align}

Since 
\begin{equation}
T_H = \Xi T_U + \frac{M_H}{2\Xi} \frac{\partial \Xi}{\partial S} \, ,
\end{equation}
it is significant that the temperature from the UTT coincides with the temperature from Hawking's proposal, that is, $T_U = T_H = T$. Moreover, new thermodynamic descriptions (with the same temperature) can be constructed using the exact isohomogeneous transformation in Eq.~\eqref{g-til}, showing that the UTT is not the only one that shares the same definition of temperature with Hawking's proposal.

It is also a feature of the UTT that the volume term deviates from the geometric volume, $V_{H}$, by the addition of an extra term [as it is in Eq.~\eqref{Dolan volume}]. The origin of this term has been revealed by the isohomogeneous transformations developed in the present work.

Another point of interest is that the Killing field that generates the UTT can be determined by the functions $g$ and $h$ from Eq.~\eqref{killing1},
\begin{equation}
\frac{1}{\Xi }K = \left[ \frac{1}{\Xi }\partial _{t} - \frac{ar_{+}^{2}}{%
L^{2} \left( a^{2} + r_{+}^{2} \right) }\partial _{\phi } \right] + 
\underbrace{ \left[ \frac{\Omega _{H}}{\Xi } + \frac{ar_{+}^{2}}{L^{2}
\left( a^{2} + r_{+}^{2} \right) } \right] } _{\Omega_U} \partial _{\phi }
\,.  \label{dolan killing}
\end{equation}
This Killing field is null and normal to the horizon, as it should be. Result~\eqref{dolan killing} emphasizes the geometric interpretation of the UTT.

Still on the geometric approach to the UTT, an observation is in order. It is commonly stated that the temperature from the UTT coincides with the black-hole surface gravity. However, this is not an accurate statement. In the geometric construction of a thermodynamic theory, the surface gravity is specified by the normalization of the Killing field normal to the horizon. While it is true to state the equality of the temperature from the UTT with the surface gravity from Hawking's proposal, notice that the Killing field associated with the UTT is normalized by a factor of $\nicefrac{1}{\Xi}$. Therefore, its surface gravity is multiplied by the same factor when compared to the surface gravity of Hawking's proposal. In other words, the temperature $T$ does not match the surface gravity of the UTT. This should be the case, since this thermodynamics is obtained from Hawking's proposal by setting $h \ne 0$.

\subsection{From Hawking's proposal to an alternative thermodynamic theory}

\label{sec:new-theory}

We can look for a theory that minimally modifies Hawking's proposal.
From Eq.~\eqref{nht}, this can be done by setting $h=0$. In this case, all thermodynamic quantities differ only by one (and the same) factor $g=\Xi^{n}$, 
\begin{equation}
h=0 \Longrightarrow T=\Xi ^{n}T_{0} \, , \,\,\, \Omega _{H}=\Xi ^{n}\Omega
_{0} \, , \,\,\, V_{H}=\Xi^{n}V_{0} \,.
\end{equation}
Moreover, from Eq.~\eqref{g-x}, it is straightforward to check that this
case comes from 
\begin{equation}
n = \frac{1}{2} \implies g = \sqrt{\Xi} \,, \,\,\, h = 0 \, , \,\,\, \xi = 
\sqrt{\Xi} \, .
\end{equation}
In this configuration, the proper thermodynamic theory associated with Hawking's is the one with energy 
\begin{equation}
M_{A} \equiv M_{0} = \frac{M_{1}}{g}=\frac{M_{H}}{\sqrt{\Xi }}=\frac{m}{\Xi
^{\frac{3}{2}}} \, .
\end{equation}
The subindex $A$ stands for alternative thermodynamic theory, abbreviated ATT.

The remaining thermodynamic variables are 
\begin{equation}
T_{A}=\frac{T}{\sqrt{\Xi }}\,,\,\,\,\Omega _{A}=\frac{\Omega _{H}}{\sqrt{\Xi 
}}\,,\,\,\,V_{A}=\frac{V_{H}}{\sqrt{\Xi }}\,.
\end{equation}
This thermodynamic version of Hawking's proposal is the one where the
temperature coincides with the surface gravity. In this case, we have $\nicefrac{\kappa_{+}}{\sqrt{\Xi}}$, where $\kappa_{+}$ is given by Eq.~\eqref{eq:kappa}. Since this property is not shared by any other
construction, this theory is the closest one to the usual thermodynamics of asymptotically flat black holes.

The formalism from the previous section guarantees that the first law is
satisfied: 
\begin{equation}
\mathrm{d}M_{A}=T_{A}\mathrm{d}S+\Omega _{A}\mathrm{d}J+V_{A}\mathrm{d}P \,.
\end{equation}
Furthermore, any thermodynamics for Kerr-anti--de Sitter black holes is
connected to the ATT by
\begin{equation}
h=M_{A} \, , \,\,\, g=g(J,P,N) \, .
\end{equation}
Therefore, the ATT is the only theory which corrects Hawking's proposal by only a multiplicative factor.

The Killing field related to the ATT is 
\begin{equation}
K=\frac{\partial _{t}}{\sqrt{\Xi }}+\frac{\Omega _{H}}{\sqrt{\Xi }}\partial
_{\phi } \,,  \label{new killing BL}
\end{equation}
which is null and normal to the horizon, as expected. Working with adapted
coordinates $(t^{\prime}, r, \theta, \phi)$, where 
\begin{equation}
t^{\prime } = \sqrt{\Xi} \, t \, ,  \label{eq-t-prime}
\end{equation}
the Killing vector field $K$ is written as 
\begin{equation}
K=\partial _{t^{\prime }}+\Omega_{A}\partial _{\phi } \, .
\label{eq-K-tprime}
\end{equation}

With the chart $(t^{\prime}, r, \theta, \phi)$, KadS metric in Eq.~\eqref{second form KadS} assumes the form 
\begin{equation}
ds^{2}=-\frac{N^{2}}{\Xi }dt^{\prime }{}^{2}+\frac{\rho ^{2}}{\Delta _{r}}%
dr^{2}+\frac{\rho ^{2}}{\Delta _{\theta }}d\theta ^{2}+\frac{\Sigma ^{2}\sin
^{2}\theta }{\rho ^{2}\Xi ^{2}}\big(d\phi -\omega ^{\prime }dt^{\prime }\big)%
^{2} \, ,
\end{equation}
with 
\begin{equation}
\omega^{\prime} = \frac{\omega }{\sqrt{\Xi }} \, .
\end{equation}
These considerations show that the angular velocity of the black hole and of
the asymptotic limit are, measured with the new time coordinate $t^{\prime}$,
\begin{equation}
\lim_{r\rightarrow r_{+}}\omega ^{\prime }=\frac{\Omega _{H}}{\sqrt{\Xi }}
=\Omega_{A} \, , \,\, \lim_{r\rightarrow \infty }\omega ^{\prime }=-\frac{a}{%
L^{2}\sqrt{\Xi }} \, .
\end{equation}
Thus, Hawking's proposal is improved (i.e. it satisfies the first law) by using the coordinate time $t^{\prime}$. As can be seen from Eq.~\eqref{eq-K-tprime}, the thermodynamics is now constructed from the Killing vectors written
with a coordinate basis, a convenient tool that is used extensively in the analysis of black-hole thermodynamics.

\section{Extension to arbitrary dimensions}

\label{sec:extensions}

\subsection{$D$-dimensional KadS spacetime and the quantum statistical relation}

The formalism developed in the context of the four-dimensional KadS geometry
can be generalized to higher-dimensional black holes. According to the results of Sec.~\ref{sec: generating}, a large set of KadS thermodynamics can be obtained by applying a (nonexact) isohomogeneous transformation to the nonthermodynamic Smarr relation proposed by Hawking. To construct a $D$-dimensional version for the ATT, presented in Sec.~\ref{sec:new-theory}, we extend Hawking's nonthermodynamic Smarr relation to higher dimensions and apply a nonexact isohomogeneous transformation to it. For this purpose, we will use the generalization to higher dimensions of the thermodynamic relations obtained in \cite{gibbons2005first}. In this extended scenario, the $D$-dimensional Kerr-anti--de Sitter spacetime models a spinning black hole characterized by $N$ independent rotation parameters $\{a_{i}\}$ with respect to the azimuthal angles $\{\varphi _{i}\}$, where 
\begin{equation}
N\equiv 
\begin{cases}
(D-1)/2 & \,\,,\,\text{odd }D \\ 
(D-2)/2 & \,\,,\,\text{even }D%
\end{cases}%
\,,
\end{equation}
and $D\geq 4$. The latitudinal angular coordinates $\mu_{i}$ satisfy the constraint 
\begin{equation}
\sum_{i=1}^{D-N-1}\mu _{i}^{2}=1\,.
\end{equation}
The $D$-dimensional KadS metric is \cite{gibbons2005first, gibbons2005general} 
\begin{align}
\mathrm{d}s^{2}=& -W\left( 1+\frac{r^{2}}{L^{2}}\right) \mathrm{d}\tau ^{2}+%
\frac{2m}{U}\left( W\mathrm{d}\tau -\sum_{i=1}^{N}\frac{a_{i}\mu _{i}^{2}%
\mathrm{d}\varphi _{i}}{\Xi _{i}}\right) ^{2}+\sum_{i=1}^{N}\frac{%
r^{2}+a_{i}^{2}}{\Xi _{i}}\mu _{i}^{2}\mathrm{d}\varphi _{i}^{2}  \notag \\
& +\frac{U}{X-2m}\mathrm{d}r^{2}+\sum_{i=1}^{D-N-1}\frac{r^{2}+a_{i}^{2}}{%
\Xi _{i}}\mathrm{d}\mu _{i}^{2}-\frac{1}{W\left( L^{2}+r^{2}\right) }\left(
\sum_{i=1}^{D-N-1}\frac{r^{2}+a_{i}^{2}}{\Xi _{i}}\mu _{i}\mathrm{d}\mu
_{i}\right) ^{2}\,.
\end{align}
To maintain consistency with the four-dimensional notation, the mass
parameter and the (negative) cosmological constant are denoted by $m$ and $\Lambda $, respectively. The parameters $\{\Xi _{i}\}$ and the functions $W$, $U$, and $X$ are given by 
\begin{equation}
\Xi _{i}\equiv 1-\frac{a_{i}^{2}}{L^{2}}\,, \,\,W\equiv \sum_{i=1}^{D-N-1}%
\frac{\mu _{i}^{2}}{\Xi _{i}}\,, \,\,U\equiv r^{D-2N-1}\sum_{i=1}^{D-N-1}%
\frac{\mu _{i}^{2}}{r^{2}+a_{i}^{2}}\prod_{j=1}^{N}(r^{2}+a_{j}^{2})\, ,
\,\,X \equiv r^{D-2N-3}\left( 1+\frac{r^{2}}{L^{2}}\right)
\prod_{i=1}^{N}\left( r^{2}+a_{i}^{2}\right) \, .
\end{equation}

The $N$ independent rotational parameters are associated with $N$ angular
momentum variables $\{ J_i \}$ \cite{caldarelli2000thermodynamics}. Thus,
considering also entropy and pressure, there are a total of $(N + 2)$
independent variables for the thermodynamic description. Isohomogeneous
transformations can be applied to the $D$-dimensional KadS scenario. In this
case, the quantities defined in Sec.~\ref{sec: smarr transf} are 
\begin{align}
\begin{split}
&r=D-3 \, , \,\, \alpha _{i}=D-2 \,, \,\, \alpha _{N+2}=-2 \,, \,\, i=1,
\cdots , N+1 \,, \\
&X_{k}=J_{k} \, , \,\, X_{N+2}=P \, , \,\, k=2,\cdots,N+1 \, , \\
&Y_{0}^{k}=\Omega_{0}^{k}\, , \,\, Y_{0}^{N+2}=V_{0} \, , \,\,
Y_{1}^{k}=\Omega_{1}^{k} \, , \,\, Y_{1}^{N+2}=V_{1} \,.
\end{split}%
\end{align}

From the Euclidean action formalism \cite{gibbons1977action}, the
thermodynamic quantities obey the so-called quantum statistical relation \cite{gibbons2005first}, 
\begin{equation}
E_{0}^{\left( D\right) }-T_{0}S-\sum_{i}\Omega _{0}^{i}J_{i}=T_{0}I_{D}\,,
\label{ea}
\end{equation}
where $I_{D}$ is the Euclidean action that, for the KadS geometry, takes the
form 
\begin{equation}
I_{D}=\frac{1}{T_{0}}\frac{A_{D-2}}{8\pi \prod_{j=1}^{N}\Xi _{j}}\left[ m-%
\frac{\left( r_{+}\right) ^{c}}{L^{2}}\prod_{i=1}^{N}\left(
r_{+}^{2}+a_{i}^{2}\right) \right] \,.  \label{id}
\end{equation}
In Eq.~\eqref{id}, $A_{D-2}$ is the volume of the unit ($D-2$)-sphere used
to construct the area $A$ of the event horizon, 
\begin{equation}
A_{D-2}=\frac{2\pi ^{\left( D-1\right) /2}}{\Gamma \left( \frac{D-1}{2}%
\right) }\,,\,\,A=\frac{A_{D-2}}{\left( r_{+}\right) ^{c}}\prod_{i=1}^{N}%
\frac{r_{+}^{2}+a_{i}^{2}}{\Xi _{i}}\,.
\end{equation}
The symbol $\Gamma$ denotes the usual gamma function, and the constant $c$
is defined as 
\begin{equation}
c\equiv 
\begin{cases}
0 & \,\,,\,\text{odd }D \\ 
1 & \,\,,\,\text{even }D%
\end{cases}%
\,.
\end{equation}

The thermodynamic quantities are given by 
\begin{eqnarray}
&&S\equiv \frac{A}{4}\,,\,\,\Omega _{0}^{i}\equiv a_{i}\frac{\left(
1+r_{+}^{2}L^{-2}\right) }{r_{+}^{2}+a_{i}^{2}}=\Omega _{H}^{i}+\frac{a_{i}}{%
L^{2}}\,,\,\,\Omega _{H}^{i}\equiv \frac{a_{i}\Xi _{i}}{r_{+}^{2}+a_{i}^{2}}%
\,,  \notag \\
&&E_{0}^{\left( D\right) }\equiv \frac{mA_{D-2}}{4\pi \Pi _{j}\Xi _{j}}%
\left( \sum_{i=1}^{N}\frac{1}{\Xi _{i}}-\frac{1-c}{2}\right)
\,,\,\,J_{i}\equiv \frac{ma_{i}A_{D-2}}{4\pi \Xi _{i}\Pi _{j}\Xi _{j}}\,.
\label{termo-gibbons}
\end{eqnarray}
These quantities satisfy the relation 
\begin{equation}
\mathrm{d}E_{0}^{\left( D\right) }=T_{0}\mathrm{d}S+\sum_{i}\Omega _{0}^{i}%
\mathrm{d}J_{i}\,,
\end{equation}
from where the temperature $T_{0}$ can be computed.

It should be noticed that, since $L$ is not considered a thermodynamic
variable in \cite{gibbons2005first}, the pressure and volume terms are
missing in this construction and quantities in Eq.~\eqref{ea} do not obey
the Smarr relation. The addition of these quantities to a proper KadS
thermodynamics is our goal in the next subsection.

\subsection{Thermodynamic volume and the Smarr relation}

To derive a Smarr relation from Eq.~\eqref{ea}, it is convenient to rewrite
the energy $E_{0}^{\left( D\right) }$ as 
\begin{equation}
E_{0}^{\left( D\right) }=\left( D-2\right) \frac{mA_{D-2}}{8\pi \Pi _{j}\Xi
_{j}}+L^{-2}\sum_{i=1}^{N}J_{i}a_{i}\,.  \label{ed}
\end{equation}
We recognize Eq.~\eqref{ed} as a thermodynamic Smarr relation by introducing 
\begin{equation}
P \equiv -\frac{\Lambda }{8\pi } =\frac{\left( D-2\right) \left( D-1\right) 
}{16\pi L^{2}}
\end{equation}
as a thermodynamic variable and defining a geometric volume \cite%
{cvetivc2011black} 
\begin{equation}
V=\frac{r_{+}}{D-1}A=\frac{(r_{+})^{c}A_{D-2}}{D-1}\prod_{i}\frac{
r_{+}^{2}+a_{i}^{2} }{\Xi _{i}}\,.
\end{equation}%
Using theses quantities, Eq.~\eqref{id} is written as 
\begin{equation}
I_{D} = \frac{1}{T_{0}}\left( \frac{mA_{D-2}}{8\pi \Pi _{j}\Xi_{j} } -\frac{2%
}{ D-2 }PV\right) \,.  \label{i}
\end{equation}

From Eqs.~\eqref{ea} and \eqref{ed}, it is straightforward to verify that
the theory now obeys the Smarr relation and the first law: 
\begin{equation}
(D-3)E_{0}^{\left( D\right) }=(D-2)T_{0}S+(D-2)\Omega
_{0}^{i}J_{i}-2V_{0}P\,,  \label{smd}
\end{equation}%
\begin{equation}
\mathrm{d}E_{0}^{\left( D\right) }=T_{0}\mathrm{d}S+\Omega _{0}^{i}\mathrm{d}%
J_{i}+V_{0}\mathrm{d}P~,
\end{equation}
with 
\begin{equation}
V_{0}=V+\frac{8\pi }{\left( D-2\right) \left( D-1\right) }%
\sum_{i=1}^{N}J_{i}a_{i}\,.  \label{eq:def-V0}
\end{equation}%
The thermodynamic volume $V_{0}$ in Eq.~\eqref{eq:def-V0}, obtained here
using the quantum statistical relation~\eqref{ea}, agrees with that obtained
in \cite{cvetivc2011black} using a different approach.

\subsection{Extension of Hawking's KadS thermodynamics to $D$ dimensions}
\label{sec:hawking-d-dimensional}

Based on the four-dimensional case, we can reinterpret Hawking's proposal as
the theory for which the Smarr relation involves the same temperature $T_{0}$
presented in Eq.~\eqref{smd}, but with the angular velocity of the black
hole $\Omega _{H}^{i}$ and the geometric volume $V$. Considering now the $D$%
-dimensional expression of Eq.~\eqref{smd}, and rearranging the terms, we
can write 
\begin{equation}
\left( D-3\right) E_{H}^{\left( D\right) }=\left( D-2\right) T_{0}S+\left(
D-2\right) \Omega _{H}^{i}J_{i}-2PV\,,  \label{sd}
\end{equation}%
where 
\begin{equation}
E_{H}^{\left( D\right) }\equiv \left( D-2\right) \frac{mA_{D-2}}{8\pi \Pi
_{j}\Xi _{j}}\,.  \label{eq:def-EH}
\end{equation}

We define the quantity $E_{H}$ as the $D$-dimensional version of Hawking's
energy. This energy gives a nonthermodynamic Smarr relation. Explicitly, 
\begin{equation}
T_{0}\mathrm{d}S+\Omega _{H}^{i}\mathrm{d}J_{i}+V\mathrm{d}P=\mathrm{d}%
E_{H}^{\left( D\right) }-\frac{E_{H}^{\left( D\right) }}{D-2}\sum_{i=1}^{N}%
\frac{1}{\Xi _{i}}\mathrm{d}\Xi _{i}\,.  \label{eq:intermediario}
\end{equation}
Following the strategy used in the four-dimensional case, it is
straightforward to obtain other thermodynamic descriptions comparing Eqs.~%
\eqref{eq:intermediario} and \eqref{fs}.

The extension to higher dimensions of Hawking's proposal presented here
differs from the five-dimensional case in \cite{hawking1999rotation}. This
is because, in the present work, the angular momentum is given by a Komar
integral in the frame in which the black hole spins with an angular velocity 
$\Omega_H^i$. Moreover, our generation satisfies a first law and the quantum statistical relation, while the one in \cite{hawking1999rotation} does not.

\subsection{The alternative thermodynamic theory in $D$ dimensions}

The ATT, presented in Sec.~\ref{sec:new-theory}, can be reinterpreted as the
theory obtained from Hawking's thermodynamics with an isohomogeneous
transformation characterized by $h=0$. Extending this idea to $D$
dimensions, 
\begin{equation}
g=\exp \left( \sum_{i=1}^{N}\frac{\ln \Xi _{i}}{D-2}\right)
=\prod_{i=1}^{N}\Xi _{i}^{\frac{1}{D-2}}\,.  \label{eq:g-d-dimensional}
\end{equation}
The function $g$ in Eq.~\eqref{eq:g-d-dimensional} transforms the $D$-dimensional version of Hawking's proposal into a $D$-dimensional generalization for the ATT.

The energy associated to the $D$-dimensional ATT is given by 
\begin{equation}
E_{A}^{\left( D\right) }\equiv \frac{E_{H}^{\left( D\right) }}{g}=\frac{D-2}{%
8\pi }mA_{D-2}\prod_{i=1}^{N}\Xi _{i}^{-\frac{D-1}{D-2}}\,.  \label{etd}
\end{equation}
The above construction gives a theory valid for arbitrary dimensions, but
different from the one presented in \cite{gibbons2005first}. In addition to
providing a proper Smarr relation, our construction can be seen as the
thermodynamic version for a generalization of Hawking's proposal. Also,
while \cite{gibbons2005first} states the failure of the Smarr-Gibbs-Duhem
relation for KadS, we have shown that this relation holds if the
cosmological constant is interpreted as a pressure term alongside with a
thermodynamic volume. Nevertheless, to our knowledge, the expression for the
energy in Eq.~\eqref{etd} is the first one to satisfy the first law of thermodynamics and the Smarr relation by construction, valid for $D\geq 4\,$.

\section{Final Remarks}
\label{remarks}

In this work, we propose a procedure for constructing thermodynamic descriptions for Kerr-anti--de Sitter black holes that are compatible with a Smarr formula. In order to consider asymptotically anti--de Sitter spacetimes, the Smarr formula is generalized from its standard version. For this, a vector volume is naturally related to geometries with a nonzero $\Lambda $. The derivation of the Smarr formula requires a Killing field $K$ normal to the horizon. However, this Killing field is not unique, since it can be multiplied by any function $g$ of the parameters $m$, $a$, and $\Lambda $ [or, equivalently by a function $g=g(S,J,P)$]. Furthermore, there are infinitely many ways to write $K$ as a combination of a timelike and a rotating Killing field. As a result, different choices for the Killing fields can alter the Komar integrals and thus the thermodynamic variables. While a Smarr formula gives the geometric side for the black-hole thermodynamics, the first law relates variations in the Komar integrals. Contrary to what might be expected, these integrals may not carry any information about the scale invariance of the system.

Our method follows from finding out which transformations can be performed in these thermodynamic variables that preserve homogeneity. We show that these isohomogeneous transformations have a geometric counterpart as a corresponding transformation in the Killing vectors. Specifically, it reduces $g$ to a homogeneous function of degree zero and restricts the viable combination between the timelike and angular Killing vectors. With these isohomogeneous transformations, and given a thermodynamics associated with a physical system, other descriptions can be obtained. Among all the possible KadS thermodynamics that can be generated with our formalism, two receive a special attention: the UTT, and Hawking's proposal. The explicit construction of a thermodynamic version of Hawking's approach in four dimensions and its generalization to arbitrary dimensions are also highlighted.

Although the usual thermodynamics is widely explored in the literature, a proper geometric construction for the theory is still undeveloped. One of our contributions in this article is to fill this gap, presenting the Killing vector associated with this theory. Moreover, our formalism clarifies why the thermodynamic and geometric volumes in the UTT do not coincide. In contrast, Hawking's proposal is not a proper thermodynamics, despite having an associated Smarr formula. Within our formalism, there is a wide set of isohomogeneous transformations that relate Hawking's proposal to proper thermodynamic theories, where the first law is satisfied. 

In addition, there is a specific theory that can be considered, in a precise sense, as the thermodynamic version of Hawking's proposal. This ATT is the closest to the standard asymptotically flat black-hole thermodynamic, since it is the one in which the temperature coincides with its own surface gravity. That is, considering a coherent normalization for the Killing fields. This claim can only be made because the present work provides a geometric construction for the theories. We show that, contrary to what has been previously asserted, the UTT has a surface gravity that does not agrees with its temperature. 

With the isohomogeneous transformations, new proposals for the thermodynamic
description of KadS black holes can now be made, furnishing different KadS thermodynamics. In anticipation of applications of the KadS thermodynamics in the context of the adS/CFT correspondence, the ATT is generalized from four to arbitrary dimensions. In the process, the extension of Hawking's proposal is also explicitly constructed. Our generalization differs from the five-dimensional case presented in Hawking's work, satisfying a first law and the quantum statistical relation.

The results of this work allow for further investigations. We anticipate a connection between the thermodynamic results presented in this study and the Hamiltonian dynamics of KadS black holes. In \cite{chrusciel2015hamiltonian}, the expression for the energy of the ATT also appears in this different context. Furthermore, we speculate that an analysis of our results from the perspective of concrete observers may provide a deeper understanding of the variety of thermodynamic descriptions for the black hole. Semiclassical treatments could also provide new insights in this area. Research along these lines is ongoing.

\begin{acknowledgments}
	
T.~L.~C. acknowledges the support of Coordena\c{c}\~ao de Aperfei\c{c}oamento de Pessoal de N\'{\i}vel Superior (CAPES) Brazil, Finance Code 001.
C.~M. is supported by Grant No.~2022/07534-0, S\~ao Paulo Research Foundation (FAPESP), Brazil.
	
\end{acknowledgments}

\end{document}